\documentclass[10pt]{article}
\usepackage[dvips]{graphicx}
\usepackage[english,russian]{babel}
\usepackage{wrapfig}
\usepackage{epsf}
\begin{document}

\renewcommand{\figurename}{Fig.}

\noindent{\small ÓÄÊ 61.80.Az, 74.25.Fy, 74.72.Bk, 02.70.Bf}\\

\renewcommand{\topfraction}{0.99}
\renewcommand{\textfraction}{0.01}
\begin{center}
\Large{\bf MINIMALISTIC HYBRID MODELS FOR THE ADSORPTION OF POLYMERS AND
PEPTIDES TO SOLID SUBSTRATES}\\
\bigskip
\large{\it Michael Bachmann and Wolfhard Janke}\\
\smallskip
\small{Institut f\"ur Theoretische Physik and Centre for
Theoretical Sciences (NTZ), 
Universit\"at Leipzig, 
04109 Leipzig, Germany\\
E-mail:\texttt{\{bachmann,janke\}@itp.uni-leipzig.de}}
\end{center}

{\small
We have performed chain-growth simulations of minimalistic hybrid
lattice models for polymers interacting with
interfaces of attractive solid substrates
in order to gain insights into the conformational transitions
of the polymers in the adsorption process. Primarily focusing on the 
dependence of the conformational behavior on temperature and solubility
we obtained pseudophase diagrams with a detailed structure of conformational
subphases. In the study of hydrophobic-polar peptides in the 
vicinity of different types of substrates, we found a noticeable 
substrate specificity of the assembly of hydrophobic domains in the 
conformations dominating the adsorption subphases.
}\\
{\small
We have performed chain-growth simulations of minimalistic hybrid
lattice models for polymers interacting with
interfaces of attractive solid substrates
in order to gain insights into the conformational transitions
of the polymers in the adsorption process. Primarily focusing on the 
dependence of the conformational behavior on temperature and solubility
we obtained pseudophase diagrams with a detailed structure of conformational
subphases. In the study of hydrophobic-polar peptides in the 
vicinity of different types of substrates, we found a noticeable 
substrate specificity of the assembly of hydrophobic domains in the 
conformations dominating the adsorption subphases.
}\\

\bigskip

\begin{center}
\bf INTRODUCTION
\end{center}
\label{secintro}
Understanding molecular self-assembly at organic-inorganic interfaces is 
essential for the design of related future nanotechnological applications, e.g., 
microscopic sensory devices in biomedicine and nanoelectronic circuits.
Recently, the enormous progress in the development of high-resolution
equipment allowed experiments which revealed quite interesting properties
of such hybrid interfaces as, e.g., the specific dependence of peptide
adhesion to the type of attractive substrates and peptide sequences~\cite{whaley1,goede1,willett1}. 
In studies of short peptides consisting of 12 amino acids, it was found, e.g.,
that the adhesion strength to a (100) silicon (Si) surface improved by a factor
of about 15 only by permuting the order of amino acids in this 
sequence. On the other hand, the adsorption strengths of the same
sequences to gallium arsenide (GaAs) with (100) orientation hardly differ~\cite{goede1}.
The reasons for this binding specificity are not yet understood and an appropriate
atomic model explaining the specific substrate--peptide cooperativity on the microscopic 
scale is still lacking. This problem is related to similar studies where the 
adsorption and docking behavior
of polymers is essential, e.g., protein--ligand
binding~\cite{ligand1}, prewetting and layering transitions in polymer solutions
as well as dewetting of polymer films~\cite{wetting1}, molecular pattern,
electrophoretic polymer deposition and growth~\cite{foo1}.

\bigskip

\begin{center}
\bf CONFORMATIONAL TRANSITIONS ACCOMPANYING HOMOPOLYMER ADHESION
\end{center}
\label{secpseudo}
\begin{figure}[t]
\centerline{\epsfxsize=13cm \epsfbox{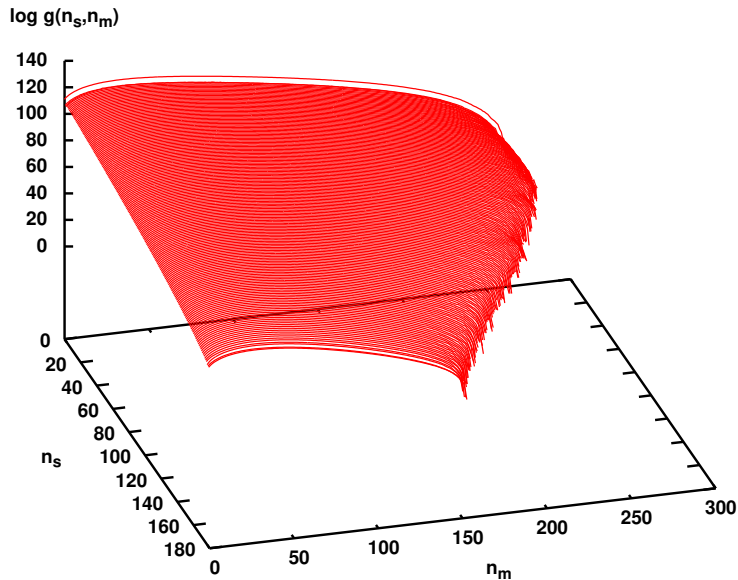}}
\caption{\label{fig:cd} 
Contact density $g(n_s,n_m)$ of a homopolymer with 179 monomers 
in a cavity with attractive 
substrate. For regularizing the number of unbound conformations (i.e., $n_s=0$), 
a steric
wall parallel to the attractive substrate was placed in a distance of $z_w=200$
lattice units.}
\end{figure}
For the following study of the conformational behavior of a homopolymer in the
adsorption process to a solid, attractive substrate, we employ a strongly 
simplified coarse-grained model. The homopolymer is modeled as interacting 
self-avoiding walk on a simple cubic (sc) lattice in implicit solvent, 
and the energy of a given conformation is 
related to the number $n_m$ of nearest-neighbor contacts of monomers being nonadjacent
along the chain. In the adsorption phase, the surface contact energy in our model
is proportional to the number of monomers in the surface-contacting layer,
and the number of monomer-substrate contacts is denoted as $n_s$. Introducing
an overall energy scale $\varepsilon_0$ (which is set to unity in the following),
and a parameter $s$ that rates the energy scales of the polymer conformation
and the surface contact energy, the minimalistic model is written 
as~\cite{vrbova1,bj1,bj2}:
\begin{equation}
\label{eq:mod}
E_s(n_s,n_m)=-\varepsilon_0 (n_s+s\,n_m).
\end{equation}   
Since the parameter $s$ effectively controls the compactness of the polymer 
conformations, it can also be interpreted as kind of solubility (the larger
the value of $s$, the worse the quality of the solvent). 
In our simulations, we applied the contact-density chain-growth method
which is a generalized variant of the multicanonical chain-growth 
algorithm~\cite{bj3}. This method allows a precise estimation of the
contact density $g(n_s,n_m)$, which is the number of conformations
with $n_s$ surface and $n_m$ intrinsic contacts. For the homopolymer
with 179 monomers used in our study, $g(n_s,n_m)$ 
ranges over more than 120 orders of magnitude, as shown in Fig.~\ref{fig:cd}.

\begin{figure}[t]
\centerline{\epsfxsize=11cm \epsfbox{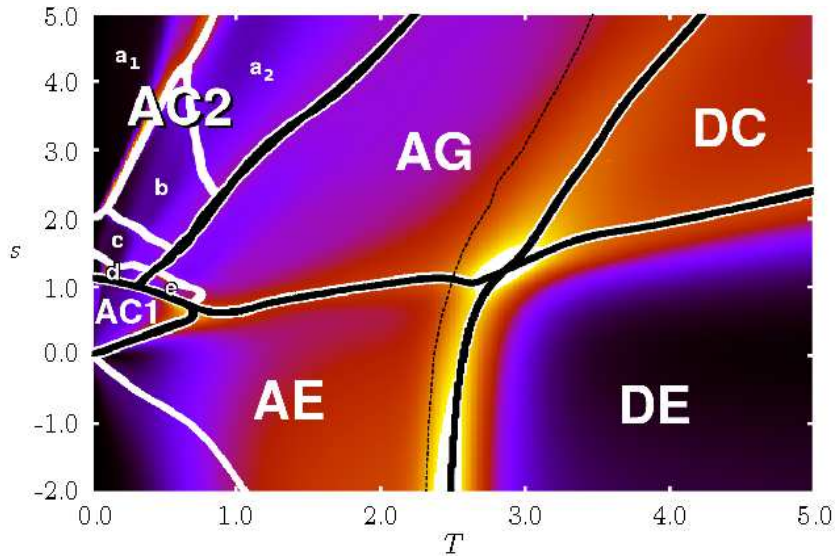}}
\caption{\label{fig:pd179} 
Profile of the specific heat as a function of temperature $T$ and
solubility $s$. Black and white lines emphasize pseudophase transitions
(cf.\ text). The dashed
black line indicates the positions, where the free-energy minima of adsorbed and
desorbed phases are degenerate, i.e., where both phases coexist with equal probability
density.}
\end{figure}
The main advantage of directly sampling $g(n_s,n_m)$ in our simulation is that 
this quantity is independent of the external parameters temperature $T$ and
solubility $s$. This means, the calculation of energetic statistical quantities
can be performed after the simulation for arbitrary values of $T$ and $s$.
Defining the partition sum by $Z=\sum_{n_s,n_m}g(n_s,n_m)\exp(-E_s/k_BT)$,
ensemble averages of functions of the contact numbers $f(n_s,n_m)$ are
obtained via 
$\langle f(n_s,n_m)\rangle=\sum_{n_s,n_m}f(n_s,n_m)g(n_s,n_m)\exp(-E_s/k_BT)/Z$.
As our main interest is focused on the conformational transitions experienced
by the polymer in the adsorption process, we consider in the following 
the specific heat,
which is here a function of temperature and solubility: 
$C_V(T,s)=(\langle E_s^2\rangle-\langle E_s\rangle^2)/k_BT^2$.
The profile of $C_V$ is shown in Fig.~\ref{fig:pd179}, where
bright regions high-light strong fluctuations, and black and white lines indicate
conformational transitions. Black lines mark transitions that are assumed to resist the
thermodynamic limit, whereas subphases specific to the precise length of the polymer
are separated by white lines. The $C_V$ profile can therefore also be interpreted
as $T$-$s$ pseudophase diagram of the hybrid system. There are two main regions, the 
desorption phases of respective compact and expanded conformations, DC and DE, and
the phases of adsorbed conformations. The adsorption regime can be divided into
compact film-like (AC1) and layered (AC2a-d) phases, globular conformations with 
surface contact (AG, AGe), and extended conformations (AE). In the low-temperature and
poor-solvent pseudophases AC1, AC2a-d and AGe, layered conformations dominate. In
AC1 a two-dimensional, maximally compact polymer film is entirely in contact with the substrate.
The less the influence of the solvent, the higher is the tendency of the polymer to
form a maximum number of intrinsic contacts by forming layers. In AC2d and AGe, double-layer
conformations dominate, in AC2c triple-layer, and in AC2b four-layer structures. The maximum 
number of intrinsic contacts in three dimensions is found in the five-layer structures in
subphases AC2a$_{1,2}$.
Note that trivial cubic symmetries are impossible as 179 is a prime number~\cite{bj1,bj2}. 

\bigskip

\begin{center}
\bf SUBSTRATE SPECIFICITY OF PEPTIDE ADSORPTION
\end{center}
\label{secfree}
\begin{figure}[t]
\centerline{\epsfxsize=13cm \epsfbox{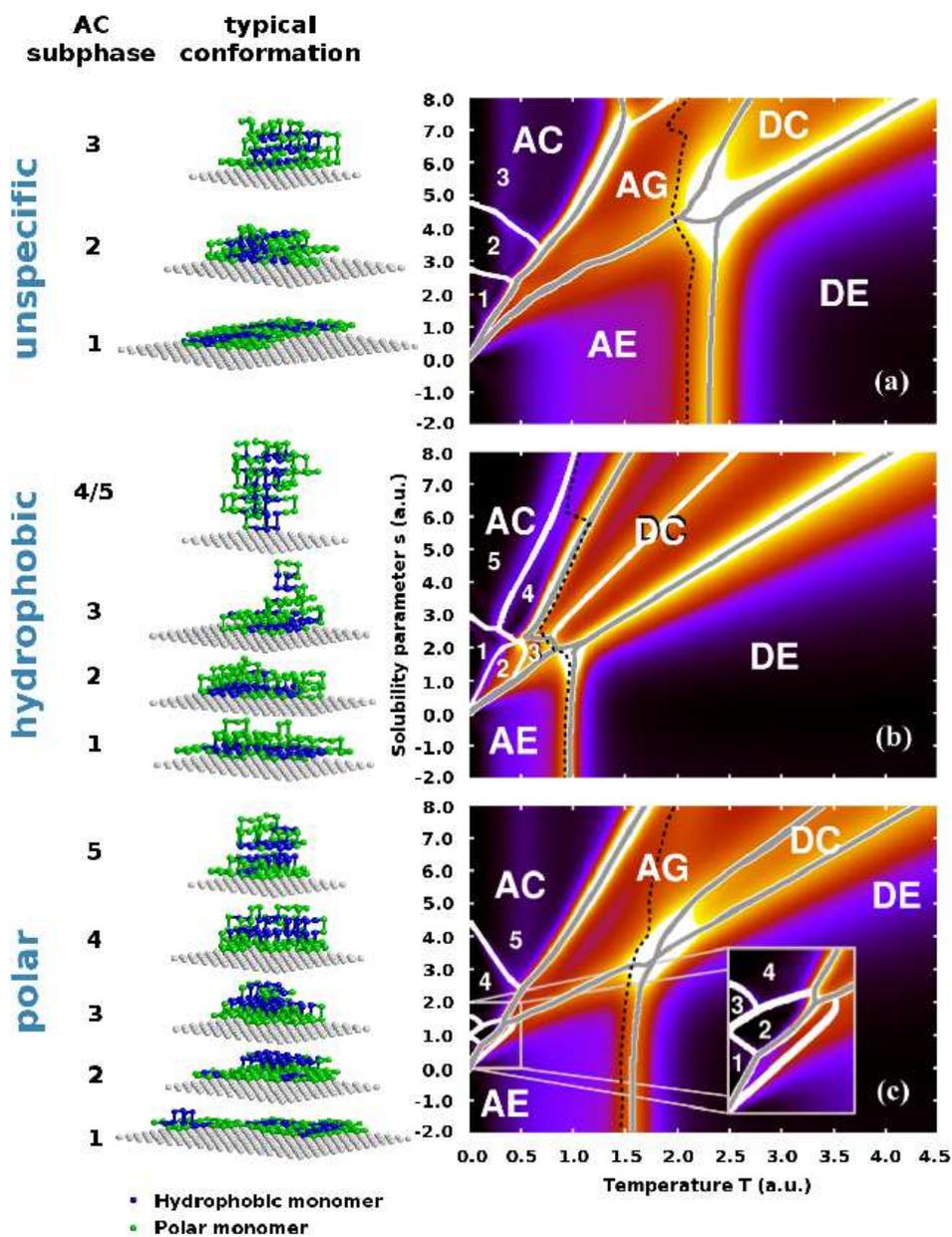}}
\caption{\label{fig:pd103} 
Specific-heat profiles for substrates being (a) unspecifically attractive, (b) hydrophobic, and (c) polar. Also shown
are typical conformations in the AC subphases.
}
\end{figure}
Considering peptides in the simplest representation, i.e., as a heteropolymer sequence of 
hydrophobic and polar residues on a simple cubic lattice, model~(\ref{eq:mod}),
with slight modifications, can also be used to study peptide adhesion to substrates.
In tertiary protein folding, the hydrophobic monomers form a compact core due to the 
hydrophobic effect, and the polar monomers screen this core from the aqueous environment.
The two types of monomers naturally allow the investigation of three types of substrates:
(a) the unspecifically attractive, (b) the hydrophobic, and (c) the polar substrate, where
the latter two are only attractive to respective hydrophobic and polar monomers with surface 
contact. Therefore, model~(\ref{eq:mod}) is substituted by~\cite{bj4}: 
\begin{equation}
\label{eq:mod2}
E_s(n_s,n_{\rm HH})=-\varepsilon_0 (n_s+s\,n_{\rm HH}),
\end{equation} 
where $n_{\rm HH}$ is the number of intrinsic nearest-neighbor contacts between nonadjacent 
hydrophobic monomers and $n_s$ symbolizes the number of attractive surface contacts and thus 
depends on the type of substrate.
In contact-density chain-growth simulations, we have studied the adsorption behavior of
an exemplified heteropolymer of 103 monomers (66 polar, 37 hydrophobic), which is the 
hydrophobic-polar transcription of the amino acid sequence of \emph{cytochrome c}.
In Fig.~\ref{fig:pd103}, the specific heat profiles of the heteropolymer in the vicinity
of the three different substrates are shown. Ridges (marked by white and gray lines) 
indicate conformational pseudophase transitions. In the bulk phases,
the typical expanded random-coil-like conformations (DE) and the compact, 
native-like folds (DC) can be distinguished. In the adsorbed regime 
we also find expanded (AE) and compact/globular (AC, AG) phases.
Typical conformations in the fine-structured AC subphases are also shown in Fig.~\ref{fig:pd103}.
The compactness of the hydrophobic domains in these subphases does not only depend on the 
solvent quality, but also on the effect how polar residues hinder the formation of 
hydrophobic domains. On the hydrophobic substrate, there is an effective steric repulsion 
of the polar monomers which are pushed off the surface layer. In the case of active attraction
of polar residues to the polar substrate, the competing tendency of the hydrophobic monomers to form 
compact, layered clusters in poor solvent leads to the loss of surface contact of these cores.
\begin{figure}[t]
\centerline{\epsfxsize=11.5cm \epsfbox{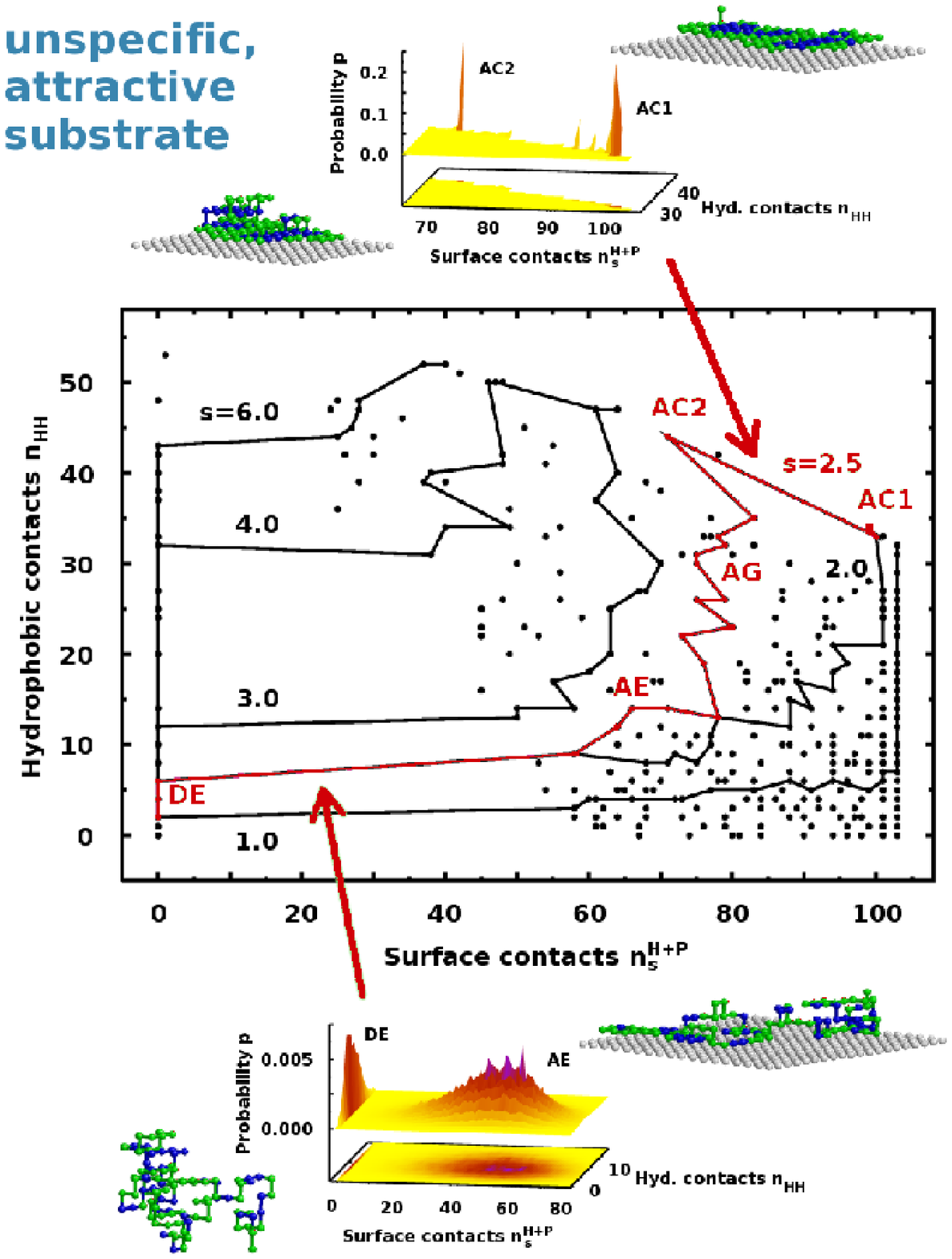}}
\caption{\label{fig:lm103} 
Contact-number map of all free-energy minima for the 103-mer
and substrate equally attractive to all monomers
in the parameter space $T\in [0,10]$, $s\in[-2,10]$. Lines illustrate contact free energy changes
with the temperature at constant solvent parameter $s$. For the exemplified
solvent with $s=2.5$, probability distributions close to the layering transition (AC1,2)
and the unbinding transition (AE to DE) are also shown.
}
\end{figure}

An alternative glance at the phase behavior is thrown by exploring the free-energy landscape
of the system. Considering the numbers of hydrophobic contacts $n_{\rm HH}$ and surface contacts
$n_s$ as natural system parameters, the free energy, expressed as function of these quantities, 
is defined as $F_{T,s}(n_s,n_{\rm HH})=E_s(n_s,n_{\rm HH})-TS(n_s,n_{\rm HH})$, where the
``microcontact'' entropy is related to the contact density $g(n_s,n_{\rm HH})$ via 
$S(n_s,n_{\rm HH})=k_B\ln\,g(n_s,n_{\rm HH})$. 
The minimum of $F_{T,s}(n_s,n_{\rm HH})$ for given external parameters $s$ and $T$ is related to 
a class of macrostates with $n_s^{(0)}$ surface and $n_{\rm HH}^{(0)}$ hydrophobic contacts which
dominates the phase. In Fig.~\ref{fig:lm103}, we have plotted the map of \emph{all} possible free-energy 
minima in the range of external parameters $T\in[0,10]$ and $s\in[-2,10]$ for the 103-mer in 
the vicinity of an unspecifically attractive substrate. 
Solid lines connect minima in the free-energy landscape when changing temperature under 
constant solvent ($s={\rm const}$) conditions. Following the exemplified trajectory for $s=2.5$
and starting at very low temperatures, it is clear from Fig.~\ref{fig:pd103}(a) that the system 
resides in pseudophase AC1, i.e., compact, film-like single-layer conformations dominate. The system
obviously prefers surface contacts at the expense of hydrophobic contacts. Increasing the temperature, 
the system experiences close to $T\approx 0.35$ a first-order-like conformational transition, and a 
second layer forms (AC2). The loss of energetically favored substrate contacts of polar monomers is 
partly compensated by the energetic gain due to the more compact hydrophobic domains. Increasing the temperature
further, globular, pancake-like conformations dominate in the globular
pseudophase AG. Reaching AE, the number of hydrophobic contacts decreases further. 
Extended, dissolved conformations dominate. The transitions from AC2 to AE via AG are comparatively 
``smooth'' (second-order-like), i.e., no immediate
changes in the contact numbers passing the transition lines are noticed. 
The situation is different when approaching the unbinding
transition line from AE  close to $T\approx 2.14$. This transition is accompanied by a dramatic loss of substrate 
contacts -- the peptide desorbs from the substrate. As the probability distribution
in Fig.~\ref{fig:lm103} shows, the unbinding transition looks again first-order-like, i.e., 
close to the transition line, there is a coexistence of adsorbed and desorbed conformations. 

\bigskip

\begin{center}
\bf CONCLUDING REMARKS
\end{center}
\label{secconc}
We have studied the conformational behavior of minimalistic hybrid interfaces of polymers and
substrates and obtained from sophisticated chain-growth simulations pseudo-phase diagrams
which exhibit a rich pseudophase structure, in particular in the adsorption regime. We could 
also show that the adsorption of heteropolymers is specifically dependent on the hydrophobic 
or polar character of the substrate. These results, whose experimental verification is still
pending, are of particular interest for future applications of hybrid organic-inorganic 
materials in nanotechnology and biomedicine.
%
%

This work is partially supported by a DFG (German Science Foundation) grant
under contract No.\ JA 483/24-1. We thank the John von Neumann Institute for 
Computing (NIC), Forschungszentrum
J\"ulich, for providing access to their 
supercomputer JUMP under grant No.\ hlz11.
\def\refname{
\end{document}